\documentclass[11pt,a4paper]{article}
\pdfoutput=1

\usepackage{amsmath, amssymb, graphics, setspace}
\usepackage{jheppub}

\usepackage[T1]{fontenc}
\usepackage[utf8]{inputenc}
\usepackage{amsfonts}
\usepackage{latexsym}
\usepackage{amsmath}
\usepackage{amssymb}
\usepackage{amssymb}
\usepackage{slashed}
\usepackage{upgreek}

\newcommand {\mathsym}[1]{{}}
\newcommand {\unicode}[1]{{}}

\usepackage{tensor}
\newcommand{\trlstens}[2]{\tensor{\tilde{#1}}{#2}}

\usepackage{slashed}
\usepackage{youngtab}

\newcommand{\beq}{\begin{equation}}
\newcommand{\eeq}{\end{equation}}

\newcommand{\beqa}{\begin{eqnarray}}
\newcommand{\eeqa}{\end{eqnarray}}

\newcommand{\bn}{\begin{equation}}
\newcommand{\en}{\end{equation}}
\newcommand{\by}{\begin{eqnarray}}
\newcommand{\ey}{\end{eqnarray}}

\newcommand{\iu}{\ensuremath{\mathrm{i}}} % Imaginary unit
\newcommand{\rd}{\ensuremath{\mathrm{d}}}

\let\Re\undefined
\let\Im\undefined
\DeclareMathOperator{\Re}{Re}
\DeclareMathOperator{\Im}{Im}

\newcommand{\Liederivative}[1]{\ensuremath{\mathcal{L}_{#1}}}
\def\SU(#1){\ensuremath{\mathrm{SU}(#1)}}
\def\Spin(#1){\ensuremath{\mathrm{Spin}(#1)}}

\newcommand{\metric}{\ensuremath{g}}
\newcommand{\KillingK}{\ensuremath{K}}

\title{Supersymmetric geometries of IIA supergravity II}

\author[a]{Ulf~Gran,}
\author[b]{George Papadopoulos,}
\author[a]{Christian von Schultz,}
\affiliation[a]{Fundamental Physics, Chalmers University of Technology\\
SE-412 96 G\"oteborg, Sweden}
\affiliation[b]{Department of Mathematics, King's College London\\
Strand, London WC2R 2LS, U.K.\vspace{1ex}}
\emailAdd{ulf.gran@chalmers.se}
\emailAdd{george.papadopoulos@kcl.ac.uk}
\emailAdd{christian.von.schultz@chalmers.se}

\abstract{We solve the Killing spinor equations of standard and massive IIA supergravities
for a Killing spinor whose isotropy subgroup in $\mathrm{Spin}(9,1)$ is $\SU(4)$ and identify the geometry of the spacetime. We demonstrate that the Killing spinor equations impose some mild
 constraints on the geometry of the spacetime which include the existence of a time-like Killing vector field which  leaves the fields and the Killing spinor invariant.}

\keywords{Supergravity Models, Superstring Vacua}

\begin{document}
\maketitle

\setcounter{page}{2}

%%%%%%%%%%%%%%%%%%%%%%%%%%%%%%%%%%%%

\section{Introduction}

The systematic investigation  of the supersymmetric backgrounds of standard \cite{giani}-\cite{huq} and massive \cite{romans} IIA supergravity
has been initiated in \cite{maxsusy} where the maximally supersymmetric backgrounds
have been classified.  Furthermore, in \cite{n31iia} it has been shown, following a similar result for IIB supergravity in \cite{n31iib},  that all backgrounds preserving 31
supersymmetries are maximally supersymmetric. More recently, it has been found in \cite{iiageomI}, using spinorial geometry \cite{spingeom}, that there
are four different types of massive (IIA) backgrounds  preserving one supersymmetry  each associated to one of the four types of non-trivial  orbits of the spin group $\mathrm{Spin}(9,1)$ on the 32-dimensional
Majorana spinor representation. Furthermore, the geometry of backgrounds
associated to orbits with isotropy groups $\mathrm{Spin}(7)$ and $\mathrm{Spin}(7)\ltimes \mathbb{R}^8$ has been
described  \cite{iiageomI}.

The main purpose of this paper is to present  the geometry of (massive) IIA backgrounds preserving one supersymmetry which are
associated to the orbit with isotropy group $\SU(4)$. The geometry of the backgrounds whose Killing spinor represents the remaining  orbit, whose isotropy group is $G_2$, will be presented elsewhere.
 A spinor representative of the orbit
with isotropy group $\SU(4)$ is
\begin{equation}
\epsilon =
    f\left(1 + e_{1234}\right)
    + g_1 \left(e_5 + e_{12345}\right)
    + \iu g_2 \left(e_5 - e_{12345}\right)
\label{su4spinor}
\end{equation}
where $f$, $g_1$ and $g_2$ are real constants with $f \neq 0$, $g_2 \neq 0$,  which later will be promoted to spacetime
functions.  Note that if $g_2=0$, then the isotropy group enhances to $\mathrm{Spin}(7)$.

The spacetime of backgrounds preserving one supersymmetry does not necessarily undergo a reduction of its  structure
group. Typically, one expects that
the Killing spinor will change orbit type from patch to patch.  As the Killing spinor solves a parallel transport equation,  it is no-where vanishing on the spacetime provided that the fields are smooth.  However, the mere existence
of a no-where vanishing section of the Spin bundle does not impose a topological restriction
on the spacetime. In particular, the G-structure of the spacetime does not necessarily reduce. This is because the  rank of the Spin bundle is much larger than the spacetime dimension and so no-where zero
sections are allowed.
Nevertheless, the spinors with the four different isotropy groups describe locally all possible geometries of backgrounds preserving one supersymmetry.
Furthermore, there are backgrounds for which the Killing spinor does not change orbit type.
The structure group of such backgrounds reduces to a subgroup of the stability group
of the corresponding orbit.

Having identified a representative spinor of the $\SU(4)$ orbit that we shall investigate, we
apply spinorial geometry \cite{spingeom} to turn the Killing spinor equations (KSEs) into a linear system for components of the fluxes and the spin connection. In this case, all these components
 are automatically expressed in representations of $\SU(4)$. The solution of the linear
system reveals the conditions on the geometry implied by the existence of a Killing spinor
as well as an expression for the fluxes in terms of the geometry. The conditions on the geometry,  given in (\ref{eqKilling}), (\ref{eqLiekappa}), (\ref{5ee}) and (\ref{5xee}),
for the existence of a Killing spinor with isotropy group $\SU(4)$ are rather mild. In particular,
it is required that the spacetime admits a time-like Killing vector field which leaves all the
fluxes invariant as well as the Killing spinor itself.  In particular, the latter condition implies that the spinorial Lie derivative
of the Killing spinor along the time-like Killing vector bilinear vanishes. The  two   additional conditions will be described later.

The expression of the fluxes in terms of the geometry is not particularly illuminating.
This is in contrast to the $\mathrm{Spin}(7)$ and $\mathrm{Spin}(7)\ltimes \mathbb{R}$ cases in \cite{iiageomI} for which
the expressions are rather simple.
Because of this, we do not solve the linear system for the fluxes completely.  Nevertheless
the vast majority of the fluxes are expressed in terms of the geometry and it is only a few
equations that are maintained in their linear system form.

Apart from the generic $\SU(4)$ backgrounds described above, there is a special class of $\SU(4)$ backgrounds for which $g_1=0$.
This special class of backgrounds is not an artefact of the gauge used to put the Killing spinor in the form (\ref{su4spinor})  as these backgrounds can also be characterised by the vanishing of
a scalar bilinear. The linear system in this special class of $\SU(4)$ models undergoes a dramatic simplification reminiscent that
of the IIB backgrounds associated with a pure spinor. The conditions on the geometry are the same as those of the generic backgrounds above. Now however,
the expression of the fluxes in terms of geometry is much simpler. We shall exploit this fact to completely solve the linear system and express the fluxes in terms of the geometry.

This paper has been organised as follows. In section two, we explain the solution of the KSEs
for generic $\SU(4)$ backgrounds and describe the geometry of spacetime. In section three,
we present the solution of the KSEs for the special class of $\SU(4)$ backgrounds with $g_1=0$.
In appendix~\ref{ap:conventions}, we describe our conventions and give some useful formulae.
In appendix~\ref{sforms}, we give the spinor form bilinears of the $\SU(4)$ invariant spinors.
In appendix~\ref{su4sol}, we present the linear system for a generic $\SU(4)$ invariant Killing spinor,
and in appendix~\ref{su4solzero} we give the solution of the linear system for the special case $g_1=0$.

%%%%%%%%%%%%%%%%%%%%%%%%%%%%%%%%%%%%

%%%%%%%%%%%%%%%%%%%%%%%%%%%%%%%%%%%%

\section{Solution of the KSEs for generic \texorpdfstring{$\SU(4)$}{SU(4)} backgrounds}

To solve the KSEs of (massive) IIA supergravity we promote $f, g_1$ and $g_2$ to real spacetime
functions and substitute the spinor $\epsilon$ in (\ref{su4spinor})  into the KSEs.%
\footnote{Throughout this paper we follow the conventions of  \cite{iiageomI} for the
fields, spinors, KSEs and field equations.}%
\textsuperscript{,}\footnote{To get standard (non-massive) IIA supergravity,
put the mass parameter $S = 0$ in what follows.} We assume that $f, g_2$ do not vanish while
$g_1$ may vanish at some points but otherwise it is generically non-zero. Then we apply the spinorial geometric technique
described in \cite{spingeom} to turn the KSEs into a linear system which contains components of the
fluxes and of the spin connection as variables. All these components are naturally expressed in
$\SU(4)$ representations and so no further work is required to re-express the linear system
in terms of the representations of the isotropy group of the Killing spinor. Then, the linear system is solved by expressing some of the components of the fluxes in terms of the spin connection.  In addition, the linear system imposes some conditions on the geometry of the spacetime.
In the context of spinorial geometry these appear as linear equations which involve
only the components of the spin connection as well as the functions $f, g_1$ and $g_2$ and the dilaton $\Phi$.
Typically these geometric conditions can be re-expressed in many different ways; one such way involves
the differentials of the fundamental forms of $\SU(4)$.

We do not present the linear system associated with  the KSEs of (massive) IIA supergravity  for the (\ref{su4spinor}) spinor in its original form as this is derived from the application
of spinorial geometry. Instead, we give a refined version which approaches the solution
of the linear system
in appendix~\ref{su4sol}, and in what follows we shall explore the properties of this solution.

\subsection{Geometry of spacetime}

To describe the geometry of spacetime, first note that from spinorial geometry considerations
there is a frame $e^A$ such
the spacetime metric can be written as
\begin{equation}
\rd s^2= 2 e^+ e^- + \rd s^2_{(8)}= -(e^0)^2 + (e^5)^2 + \rd s^2_{(8)}~,
\end{equation}
where $\rd s^2_{(8)}=2 \delta_{\alpha\bar\beta} e^\alpha e^{\bar\beta}$ is a metric transverse to the lightcone directions $e^+$ and $e^-$ and $\alpha, \beta=1,\dots, 4$ are holomorphic $\SU(4)$ frame indices.
Of course such a splitting is only on the tangent bundle of the spacetime and the metric potentially
depends on all spacetime coordinates.

Next let us consider the form bilinears of the Killing
spinor $\epsilon$ in (\ref{su4spinor}).
These are given in appendix~\ref{sforms}\@. Before we proceed
observe that under a boost in the $5$ direction, which is a gauge symmetry of the theory,
$f\rightarrow \ell f$ while $g_1, g_2\rightarrow \ell^{-1} g_1, \ell^{-1} g_2$. As a result,
we can choose the gauge $f^2=g_1^2+ g_2^2$.
In this gauge, one finds that the 1-form bilinears become
\begin{equation}
\KillingK = f^2 e^0~,\quad X = f^2 e^5~,
\label{kvector}
\end{equation}
after an additional trivial numerical normalization.

It now turns out that the geometric conditions \eqref{su4sol:part0logf}, \eqref{su4sol:part5logf},
\eqref{su4sol:Omega505}, \eqref{su4sol:Omega0trre},
\eqref{su4sol:Omega50ho}, \eqref{su4sol:Liedf},
\eqref{su4sol:Omegasymhoho0} and \eqref{su4sol:Omegasymhoaho0}  imply that $\KillingK$
is associated with a Killing vector field, i.e.
\begin{equation}
\nabla_{(A} \KillingK_{B)} = 0~.
\label{eqKilling}
\end{equation}
In fact, this vector field leaves invariant all the other fields of the theory. This was
expected as it is well known that this 1-form bilinear generates a symmetry of the theory.  In addition, the geometric
conditions imply that
\begin{equation}
[X,K]=0~.
\end{equation}
This is significant  even though $X$ does not generate a symmetry for the backgrounds, as it is possible to adapt
local independent coordinates for both $K$ and $X$ so that the expression for the fields can be simplified. For example, if $K=\partial_\tau$
and $X=\partial_\sigma$, the spacetime metric can be written as
\begin{equation}
\rd s^2=- f^{4} (\rd\tau+ m)^2+ f^{4} (\rd\sigma+n)^2 + \rd s^2_{(8)}~,
\end{equation}
where $m$ and $n$ are 1-forms and  $\rd s^2_{(8)}$ is a metric in the directions transverse to $X$ and $K$. All the components of the
metric do not dependent  on $\tau$ but they can depend on $\sigma$ and all the remaining coordinates of the spacetime.

Furthermore, one can verify that the geometric conditions \eqref{su4sol:part0logf}, \eqref{su4sol:part5logf},
\eqref{su4sol:part0logg}, \eqref{su4sol:Omega0trim},
\eqref{su4sol:LieOmega}, \eqref{su4sol:Liedf} and \eqref{su4sol:ahoahoOmega0}
can be expressed as

\begin{equation}
\Liederivative{\KillingK}\epsilon = 0~,
\label{eqLiekappa}
\end{equation}
where $\Liederivative{\KillingK}$ is the spinorial Lie derivative associated with
the Killing vector field $\KillingK$, i.e.
$ \Liederivative{\KillingK} = \KillingK^\mu\,\nabla_\mu
  + \tfrac{1}{4}\,\nabla_\mu\,\KillingK_\nu\,\Gamma^{\mu\nu}$.
Therefore, the Killing spinor $\epsilon$ is invariant under the motion generated
by $\KillingK$.

The two  conditions above provide a geometric description of all conditions imposed
by the KSEs on the geometry of spacetime apart from
\eqref{su4sol:part5logg} and \eqref{su4sol:part5phi}. For the former one can show, after some calculation, that this condition
can be expressed as
\begin{equation}
\partial_5\log\frac{g}{\bar g}=\frac{\iu}{3!} \Liederivative{5} (\mathrm{Re}\, \epsilon_{A_1 A_2A_3A_4})\,
 \mathrm{Im}\, \epsilon^{A_1 A_2A_3A_4}~,
 \label{5ee}
 \end{equation}
 where
 $g \equiv g_1 + \iu g_2$.
Similarly, \eqref{su4sol:part5phi} can be written as
\begin{equation}
\partial_5\Phi=\partial_5\log(g\bar g)-\frac{1}{6} \mathrm{Re}\,( \epsilon^{\gamma_1\gamma_2\gamma_3\gamma_4}
\nabla_{\gamma_1} \epsilon_{5\gamma_2\gamma_3\gamma_4})~.
\label{5xee}
\end{equation}
This concludes the discussion of the geometry.

\subsection{Fluxes in terms of geometry}

The solution of the linear system also expresses some of the components of the fluxes in terms of the
geometry. As these components are already in $\SU(4)$ representations, the solution of the linear system
presented in appendix~\ref{su4sol} is already in the required form.  In particular denoting collectively
the 2-form, 3-form and 4-form field strengths with $G^k$, $k=2,3,4$, we first decompose them as
\begin{eqnarray}
G^k&=& e^0\wedge e^5\wedge G^k_{(k-2)}+ e^0\wedge G^k_{0(k-1)}+ e^5\wedge G^k_{5(k-1)}+ G^k_{(k)}
\cr &=&
e^+\wedge e^-\wedge G^k_{(k-2)}+ e^-\wedge G^k_{-(k-1)}+ e^+\wedge G^k_{+(k-1)}+ G^k_{(k)}~,
\label{dec}
\end{eqnarray}
where the  subscript in the brackets denotes the degree of the form in the directions transverse
to $e^0$ and $e^5$ or equivalently $e^+$ and $e^-$.  Furthermore each component $G^k_{(k-2)}$, $G^k_{0(k-1)}$, $G^k_{5(k-1)}$ and $G^k_{(k)}$, or equivalently  $G^k_{(k-2)}$, $G^k_{-(k-1)}$, $G^k_{+(k-1)}$ and $G^k_{(k)}$,
is decomposed further in $SU(4)\subset SO(8)$ representations. The resulting components are given
it terms of the geometry as in appendix~\ref{su4sol}. It may appear that the corresponding expressions are not covariant as they contain components of the spin connection. However this is not the case. All the components of the spin connection that appear in the expressions for the fluxes in appendix~\ref{su4sol} actually transform
as tensors under $\SU(4)$ gauge transformations. So provided that the spacetime has an $\SU(4)$ structure these expressions  patch in a covariant manner on the spacetime manifold $M$.

For example the $(1,2)$ and traceless part of $G^4_{-(3)}$ and $G^4_{+(3)}$ are given in (\ref{g478})
and (\ref{g479}), respectively, in terms of the geometry and the (1,2) and traceless component of
$H^3_{(3)}$. Note that this  component of $H^3_{(3)}$
is not restricted by the KSEs. In a similar way one can read the remaining equations in appendix~\ref{su4sol}\@.
However, solving the equations involving the $P^-$ and $P^+$ projections in appendix~\ref{su4sol}, although possible, does not give an illuminating answer.  As a result, we shall not attempt to give the full expression of the
fluxes in terms of the geometry. The form already presented in appendix~\ref{su4sol} is more economical.

%%%%%%%%%%%%%%%%%%%%%%%%%%%%%%%%%%%%

\section{Solution of the KSEs for a special case of \texorpdfstring{$\SU(4)$}{SU(4)} backgrounds}

The special backgrounds that we shall be considering  are those for which the scalar bilinear $\sigma$ in appendix~\ref{sforms} vanishes.
As $f \neq 0$, this condition implies that $g_1 = 0$.  As the vanishing of a scalar
is a covariant statement, it can be imposed globally on a manifold because it is consistent
with the patching conditions. So this special class of backgrounds does not depend on the
 choice of representative for the spinor $\epsilon$ in (\ref{su4spinor}).

These special IIA backgrounds are reminiscent of the IIB backgrounds in \cite{iibsu4} which admit a pure spinor
as Killing spinor. In particular, the linear system, as in the IIB case,  simplifies considerably.
This particularly applies in the gauge $g \bar g=f^2$ that we shall use throughout. As a result
$f^2=g_2^2$ and so $f=\pm g_2$. In what follows we shall take $f=g_2$; the other case can be treated
symmetrically.

\subsection{Geometry of spacetime}

The geometry of spacetime is as that described in the generic case after imposing $g_1=0$ on all
geometric conditions in section 2.1. The only significant change is in equation (\ref{5ee})
where now $\partial_5 \log(g/\bar g)=0$.

\subsection{Fluxes in terms of geometry}

The expression of the fluxes in terms of the geometry is somewhat simpler in this
case. The linear system can be solved and the solution has been presented in appendix~\ref{su4solzero}\@.

It can be seen from the solution that not all components of the fluxes are given in terms of the
geometry. For example the traceless (1,1) component of $G_{+-{(2)}}^4$ and the traceless
(2,2) component of $G^4_{(4)}$ are not given in terms of the geometry and so they are not constrained
by the KSEs. The components
of the fluxes that are given in terms of the geometry have been expressed in terms of components
of the spin connection as well as in terms of components of the fluxes that are not constrained.
There are various ways to  re-express these fluxes,  for example in terms of the spinor
form bilinears given in appendix~\ref{sforms}\@.
For this one has to use the relation of these to the components of the spin connection as given in appendix~\ref{ap:conventions}\@.

\acknowledgments
UG is supported by the Knut and Alice Wallenberg Foundation. GP is partially supported by the STFC grant ST/J002798/1.

\appendix

\section{Conventions and useful \texorpdfstring{$\SU(4)$}{SU(4)} formulae}
\label{ap:conventions}

The conventions for the fields, KSEs and field equations of standard \cite{giani}-\cite{huq} and massive \cite{romans} IIA
 supergravity that we use are as those in \cite{iiageomI} which have been adapted from the string frame formulation of the theory as in \cite{e1} and \cite{e2}.

Our spinor conventions and in particular the null basis we use to express the spinor representative
(\ref{su4spinor})
of the $\SU(4)$ orbit are as in \cite{spingeom} and \cite{msyst}. In these works, it is also explained how to realise the spinors in terms of forms.

The components of the spin connection that appear in the linear system that arises
from the solution of the KSEs in the context of spinorial geometry can be expressed
in terms of tensors on the spacetime and its covariant derivatives. In particular, we have that
the Nijenhuis tensor is given by
\bn
{\cal N}^\alpha{}_{\bar \beta \bar \gamma} = 8 \Omega_{[\bar \beta,\bar \gamma]}{}^\alpha~,
\en
and the Lee forms of the Hermitian form $\omega$ and of the  (4,0) fundamental $\SU(4)$ form
$\chi=4\epsilon$ are given by
\by
(\theta_\omega)_\alpha &=& 2 \Omega^\gamma{}_{,\gamma \alpha} ~, \\
(\theta_{\Re \chi})_\alpha &=& \Omega^\gamma{}_{,\gamma \alpha} -\Omega_{\alpha , \gamma}{}^\gamma ~,
\ey
respectively.
The covariant derivative of the Hermitian form is
\by
\nabla_A \omega_{\beta \bar \gamma} &=& 0 ~, \\
\nabla_A \omega_{\beta \gamma} &=& 2 \iu \Omega_{A,\beta \gamma} ~,
\ey
where $\omega_{\alpha\bar \beta}= -\iu g_{\alpha \bar \beta}$. Furthermore, we have
\let\oldindexmarker\indexmarker%
\renewcommand\indexmarker{\phantom{\gamma}}%
\by%
% 4 contractions:
\epsilon^{\gamma_1\gamma_2\gamma_3\gamma_4}\,
    \nabla\!_A\,
    \epsilon_{\gamma_1\gamma_2\gamma_3\gamma_4}
    &=& 24\,\tensor{\Omega}{_A_,_\gamma^\gamma} ~,\\
\epsilon^{\gamma_1\gamma_2\gamma_3\gamma_4}\,
    \nabla_{\gamma_1}\,
    \epsilon_{a\gamma_2\gamma_3\gamma_4}
    &=& 6\,\tensor{\Omega}{_\gamma_,_a^\gamma}~,\\
\epsilon^{\gamma_1\gamma_2\gamma_3\gamma_4}\,
    \nabla_{\gamma_1}\,
    \epsilon_{\bar\beta\gamma_2\gamma_3\gamma_4}
    &=& 6\,\tensor{\Omega}{_\gamma_,_{\bar\beta}^\gamma}~,\\
\epsilon^{\gamma_1\gamma_2\gamma_3\gamma_4}\,
    \nabla_{\gamma_1}\,
    \epsilon_{\beta\gamma_2\gamma_3\gamma_4}
    &=& 6\,\tensor{\Omega}{_\beta_,_\gamma^\gamma}~,\\
%\ey\by%
% 3 contractions:
\tensor{\epsilon}{_{\bar\alpha}^{\gamma_1}^{\gamma_2}^{\gamma_3}}\,
    \nabla\!_A\,
    \epsilon_{\beta\gamma_1\gamma_2\gamma_3}
    &=& 6\,\tensor{\metric}{_{\bar\alpha}_\beta}\,\tensor{\Omega}{_A_,_\gamma^\gamma}~,\\
\tensor{\epsilon}{_{\bar\alpha}^{\gamma_1}^{\gamma_2}^{\gamma_3}}\,
    \nabla\!_A\,
    \epsilon_{\bar\beta\gamma_1\gamma_2\gamma_3}
    &=& 6\,\tensor{\Omega}{_A_,_{\bar\beta}_{\bar\alpha}}~,\\
\tensor{\epsilon}{_{\bar\alpha}^{\gamma_1}^{\gamma_2}^{\gamma_3}}\,
    \nabla\!_A\,
    \epsilon_{a\gamma_1\gamma_2\gamma_3}
    &=& 6\,\tensor{\Omega}{_A_,_a_{\bar\alpha}}~, \\
\tensor{\epsilon}{_{\bar\alpha}^{\gamma_1}^{\gamma_2}^{\gamma_3}}\,
    \nabla_{\gamma_1}\,
    \epsilon_{a\beta_1\gamma_2\gamma_3}
    &=& 2\,\tensor{\Omega}{_\beta_,_a_{\bar\alpha}}
        - 2\,\tensor{\metric}{_{\bar\alpha}_\beta}\,
            \tensor{\Omega}{_\gamma_,_a^\gamma}~,\\
\tensor{\epsilon}{_{\bar\alpha}^{\gamma_1}^{\gamma_2}^{\gamma_3}}\,
    \nabla_{\gamma_1}\,
    \epsilon_{\bar\beta_0\beta_1\gamma_2\gamma_3}
    &=& 2\,\tensor{\Omega}{_{\beta_1}_,_{\bar\beta_0}_{\bar\alpha}}
        - 2\,\tensor{\metric}{_{\bar\alpha}_{\beta_1}}\,
            \tensor{\Omega}{_\gamma_,_{\bar\beta_0}^\gamma}~,\\
\tensor{\epsilon}{_{\bar\alpha}^{\gamma_1}^{\gamma_2}^{\gamma_3}}\,
    \nabla_{\gamma_1}\,
    \epsilon_{\beta_0\beta_1\gamma_2\gamma_3}
    &=& 4\,\tensor{\metric}{_{\bar\alpha}_[_{\beta_0}}\,
        \tensor{\Omega}{_{\beta_1}_]_,_\gamma^\gamma}~,
\ey
where $a \in \{0, 5\}$. All these formulae can be used to express the solutions of the linear system in terms of the spinor bilinears. These formulae have also been used to describe the
geometry of the supersymmetric backgrounds.

\let\indexmarker\oldindexmarker%

\section{Spacetime forms from spinor bilinears}
\label{sforms}

To compute the form spinor bilinears of $\epsilon$ in (\ref{su4spinor}), first note that we can define another spinor $\tilde \epsilon= \Gamma_{11}\epsilon$, which need not be Killing, but is nevertheless defined on the spacetime. Using the Dirac inner product and after an appropriate normalisation of $\epsilon$,   we find a 0-form
\bn
\sigma(\epsilon,\tilde\epsilon) = - 2 f g_1~,
\en
two 1-forms
\by
\kappa(\epsilon,\epsilon) &=& f^2 (e^0 -e^5) + (g_1^2+g_2^2) (e^0 + e^5)~,\notag\\
\kappa(\epsilon,\tilde\epsilon) &=& -f^2 (e^0 -e^5) + (g_1^2+g_2^2) (e^0 + e^5)~,
\ey
a 2-form
\bn
\omega(\epsilon,\epsilon) = 2 f g_1  e^0 \wedge e^5 - 2 f g_2 \omega~,
\en
a 4-form
\bn
\zeta(\epsilon,\tilde\epsilon) = 2 f g_2 e^0 \wedge e^5 \wedge \omega -2 f \Re((g_1 + \iu g_2)\chi)+f g_1 \omega \wedge \omega~,
\en
and two 5-forms
\by
\tau(\epsilon,\epsilon) &=& f^2 (e^0 -e^5)\wedge \Re \chi + (g_1^2-g_2^2) (e^0+e^5)\wedge \Re  \chi -2 g_1 g_2 (e^0+e^5)\wedge \Im  \chi \nonumber\\
&& -\frac{1}{2} f^2 (e^0-e^5) \wedge \omega\wedge\omega -\frac{1}{2}( g_1^2+g_2^2) (e^0+e^5) \wedge \omega\wedge\omega~,\\
\tau(\epsilon,\tilde\epsilon) &=& -f^2 (e^0 -e^5)\wedge \Re \chi + (g_1^2-g_2^2) (e^0+e^5)\wedge \Re  \chi -2 g_1 g_2 (e^0+e^5)\wedge \Im  \chi \nonumber\\
&& +\frac{1}{2} f^2 (e^0-e^5) \wedge \omega\wedge\omega -\frac{1}{2}( g_1^2+g_2^2)  (e^0+e^5) \wedge \omega\wedge\omega~,
\ey
where $\omega= - \iu \sum_\alpha e^\alpha\wedge e^{\bar \alpha}$ is the Hermitian form and $\chi=4 e^1\wedge e^2\wedge e^3\wedge e^4$ is the (4,0)-form both of which are the fundamental forms of $\SU(4)$.

\section{The linear system for generic \texorpdfstring{$\SU(4)$}{SU(4)} backgrounds}
\label{su4sol}

Here we present a refined form of the linear system associated with the KSEs for
generic $\SU(4)$ backgrounds, i.e.~$g_1\not=0$. This arises after substituting \eqref{su4spinor}
into the KSEs and after some further re-arrangements of the resulting equations. The expressions presented are
nearly in the form of a solution to the linear system. The formulae are organised according to representations of $\SU(4)$. Moreover, we impose the gauge $f^2=g \bar g$ throughout.

\subsection*{Scalar representation}
The conditions that arise in the scalar representation of $\SU(4)$ are
\by
\partial_+ \log f &=& \partial_- \log f~,\\
\partial_+ \log f &=& \tfrac{1}{2}\Omega_{+,-+}~,\\
\Omega_{-,-+} &=& -\Omega_{+,-+}~,\\
\partial_+ \log g &=& \partial_- \log g~,\\
\partial_+ \log g &=&
    -\left(
        \tensor{\Omega}{_+_,_\gamma^\gamma}
        +\tfrac{1}{2}\tensor{\Omega}{^\gamma_,_+_\gamma}
        -\tfrac{1}{2}\tensor{\Omega}{_\gamma_,_+^\gamma}
    \right)
    +\tfrac{1}{2}\Omega_{+,-+}~,\\
\tensor{\Omega}{_-_,_\gamma^\gamma}
    +\tfrac{1}{2}\tensor{\Omega}{^\gamma_,_-_\gamma}
    -\tfrac{1}{2}\tensor{\Omega}{_\gamma_,_-^\gamma} &=&
       \tensor{\Omega}{_+_,_\gamma^\gamma}
        +\tfrac{1}{2}\tensor{\Omega}{^\gamma_,_+_\gamma}
        -\tfrac{1}{2}\tensor{\Omega}{_\gamma_,_+^\gamma}~,\\
        \rd \Phi_+ &=& \rd \Phi_-~,\\
\rd \Phi_+ &=& -\tfrac{1}{2}\left(
        \tensor{\Omega}{^\gamma_,_+_\gamma}
        +\tensor{\Omega}{_\gamma_,_+^\gamma}
    \right) + \Omega_{+,-+}~,\\
\tensor{\Omega}{^\gamma_,_-_\gamma}
        +\tensor{\Omega}{_\gamma_,_-^\gamma} &=&
    \tensor{\Omega}{^\gamma_,_+_\gamma}
        +\tensor{\Omega}{_\gamma_,_+^\gamma}~.
\ey
In addition the following equations restrict components of the $H$ field strength
\by
H_{-\gamma}{}^\gamma &=& 2\Omega_{-,\gamma}{}^\gamma ~,\\
H_{+\gamma}{}^\gamma &=& 2 (\Omega^\gamma{}_{,+\gamma}-\Omega_{\gamma,+}{}^\gamma+ \Omega_{+,\gamma}{}^\gamma) ~.
\ey
There are also some additional conditions which restrict the remaining fluxes
\by
\frac{1}{4!}( g \epsilon^{\bar\gamma_1 \bar\gamma_2 \bar\gamma_3 \bar\gamma_4}G_{\bar\gamma_1 \bar\gamma_2 \bar\gamma_3 \bar\gamma_4}-\bar g \epsilon^{\gamma_1 \gamma_2 \gamma_3 \gamma_4}G_{\gamma_1 \gamma_2 \gamma_3 \gamma_4}) - \sqrt{2} f (\Omega_{\gamma,+}{}^\gamma-\Omega^\gamma{}_{,-\gamma})&=&0~,
\\
g (\rd \Phi_- +\Omega_{\gamma,-}{}^\gamma) +\frac{f}{4\sqrt{2}} (F_{-+} -F_\gamma{}^\gamma - G_{-+\gamma}{}^\gamma + \frac{1}{2}G_{\gamma}{}^\gamma{}_\delta{}^\delta &&
\\
 +\frac{1}{6}\epsilon^{\gamma_1\gamma_2\gamma_3\gamma_4}G_{\gamma_1\gamma_2\gamma_3\gamma_4}+S)&=&0~,
 \notag\\
\frac{f}{\sqrt{2}} (2F_{-+}+ F_\gamma{}^\gamma-G_{-+\gamma}{}^\gamma-2 S) +2  g (\Omega_{\gamma,-}{}^\gamma-\Omega_{-,\gamma}{}^\gamma)&=& 0  ~.\ey

It is convenient for the investigation of geometry to rewrite the first set of condition in the
$05$ frame. One then finds
\by
\partial_0 \log f &=& 0
    \label{su4sol:part0logf}~,\\
\partial_5 \log f &=& -\tfrac{1}{2}\,\Omega_{0,05}
    \label{su4sol:part5logf}~,\\
\partial_0 \log g &=& 0
    \label{su4sol:part0logg}~,\\
\partial_5 \log g &=&
    -\tfrac{1}{2}\,\Omega_{0,05}
    -\left(
        \tensor{\Omega}{_5_,_\gamma^\gamma}
        +\tfrac{1}{2}\tensor{\Omega}{^\gamma_,_5_\gamma}
        -\tfrac{1}{2}\tensor{\Omega}{_\gamma_,_5^\gamma}
    \right)\label{su4sol:part5logg}~,\\
\rd \Phi_0 &=& 0
    \label{su4sol:part0phi}~,\\
\rd \Phi_5 &=&
    -\Omega_{0,05}
    -\tfrac{1}{2}\tensor{\Omega}{^\gamma_,_5_\gamma}
    -\tfrac{1}{2}\tensor{\Omega}{_\gamma_,_5^\gamma}
    \label{su4sol:part5phi}~,\\
\Omega_{5,05} &=& 0
    \label{su4sol:Omega505}~,\\
\tensor{\Omega}{_0_,_\gamma^\gamma}
    +\tfrac{1}{2}\tensor{\Omega}{^\gamma_,_0_\gamma}
    -\tfrac{1}{2}\tensor{\Omega}{_\gamma_,_0^\gamma} &=& 0
    \label{su4sol:Omega0trim}~,\\
\tensor{\Omega}{^\gamma_,_0_\gamma}
    +\tensor{\Omega}{_\gamma_,_0^\gamma} &=& 0 ~.
    \label{su4sol:Omega0trre}
\ey
Note also that the conditions on the $H$ flux can be written as
\by
\tensor{H}{_0_\gamma^\gamma} &=&
    \tensor{\Omega}{^\gamma_,_5_\gamma}
    -\tensor{\Omega}{_\gamma_,_5^\gamma}
    \label{su4sol:H0tr}~,\\
\tensor{H}{_0_\gamma^\gamma} - 2\tensor{\Omega}{_0_,_\gamma^\gamma} &=&
    \tensor{H}{_5_\gamma^\gamma}
    -2\tensor{\Omega}{_5_,_\gamma^\gamma}~.
    \label{su4sol:H5tr}
\ey
This concludes the description of the conditions in  the scalar representation.
\subsection*{Fundamental representation}
The conditions that lie in the fundamental representation of $\SU(4)$ are
\by
H_{-+\alpha} &=& 2 \Omega_{-,+\alpha} ~,  \\
\Omega_{+,-\alpha} &=& \Omega_{-,+\alpha} ~, \\
\Omega_{\alpha,-+} &=& -\frac{1}{2}\Omega_{-,-\alpha} + \frac{1}{2} \Omega_{+,+\alpha} ~, \\
\partial_\alpha \log f &=& \frac{1}{2} \partial_\alpha \log g \bar g = \frac{1}{4}H_{-+\alpha} -\frac{1}{2}\Omega_{\alpha,-+}-\frac{1}{2}\Omega_{-,-\alpha} ~, \\
\partial_\alpha \log \frac{g}{\bar g} &=& \frac{1}{6} \left(\frac{g}{\bar g}+1\right)\epsilon_\alpha{}^{\bar\gamma_1 \bar\gamma_2 \bar\gamma_3}H_{\bar\gamma_1 \bar\gamma_2 \bar\gamma_3} +\left(\frac{g}{\bar g} -1\right) \epsilon_\alpha{}^{\bar\gamma_1 \bar\gamma_2 \bar\gamma_3}\Omega_{\bar\gamma_1, \bar\gamma_2 \bar\gamma_3} \notag \\&& - H_{\alpha\gamma}{}^\gamma
-2 \Omega_{\alpha,- +} ~,\\
\partial_\alpha \Phi &=& \frac{1}{2} H_{-+\alpha} +\frac{1}{4} H_{\alpha\gamma}{}^\gamma -\frac{1}{12}\epsilon_\alpha{}^{\bar\gamma_1 \bar\gamma_2 \bar\gamma_3}H_{\bar\gamma_1 \bar\gamma_2 \bar\gamma_3}-\frac{1}{2}\Omega_{-,-\alpha} \notag\\&&-\frac{1}{2} \Omega_{\alpha,\gamma}{}^\gamma
 + \Omega^\gamma{}_{,\gamma \alpha}  +\frac{1}{2}\epsilon_\alpha{}^{\bar\gamma_1 \bar\gamma_2 \bar\gamma_3}\Omega_{\bar\gamma_1, \bar\gamma_2 \bar\gamma_3} ~,\\
g(G_{-\alpha} + G_{-\alpha\gamma}{}^\gamma) &=& \frac{f}{\sqrt{2}} (H_{\alpha\gamma}{}^\gamma -2 \Omega_{\alpha,\gamma}{}^\gamma +2 \Omega_{-,-\alpha}) ~,  \\
\frac{g}{3}\epsilon_\alpha{}^{\bar\gamma_1 \bar\gamma_2 \bar\gamma_3}G_{-\bar\gamma_1 \bar\gamma_2 \bar\gamma_3} &=& \frac{f}{\sqrt{2}} (H_{\alpha\gamma}{}^\gamma -2 \Omega_{\alpha,\gamma}{}^\gamma -2 \Omega_{-,-\alpha}) ~, \\
\frac{f}{2} (3 G_{-\alpha} - G_{-\alpha\gamma}{}^\gamma) &=& \frac{g}{\sqrt{2}} (H_{\alpha\gamma}{}^\gamma - \frac{1}{2} \epsilon_\alpha{}^{\bar\gamma_1 \bar\gamma_2 \bar\gamma_3} H_{\bar\gamma_1 \bar\gamma_2 \bar\gamma_3}+ 2 \Omega^{ \gamma}{}_{,\gamma\alpha} \notag\\
&& \qquad+ \epsilon_\alpha{}^{\bar\gamma_1 \bar\gamma_2 \bar\gamma_3}\Omega_{\bar\gamma_1, \bar\gamma_2 \bar\gamma_3}) ~,\\
\frac{f}{2} (3 G_{+\alpha} + G_{+\alpha\gamma}{}^\gamma) &=& \frac{1}{\sqrt{2}} (\bar g H_{\alpha\gamma}{}^\gamma - \frac{g}{2} \epsilon_\alpha{}^{\bar\gamma_1 \bar\gamma_2 \bar\gamma_3} H_{\bar\gamma_1 \bar\gamma_2 \bar\gamma_3}- 2\bar g \Omega^{ \gamma}{}_{,\gamma\alpha}\notag\\
&& \qquad -g \epsilon_\alpha{}^{\bar\gamma_1 \bar\gamma_2 \bar\gamma_3}\Omega_{\bar\gamma_1, \bar\gamma_2 \bar\gamma_3}) ~, \\
\frac{g}{3} \epsilon_\alpha{}^{\bar\gamma_1 \bar\gamma_2 \bar\gamma_3}G_{+\bar\gamma_1 \bar\gamma_2 \bar\gamma_3}  &=& - \bar g (G_{+\alpha} - G_{+\alpha \gamma}{}^\gamma) - 2 \sqrt{2} f \Omega_{+,+\alpha} ~,\\
\frac{f}{6\sqrt{2}} \epsilon_\alpha{}^{\bar\gamma_1 \bar\gamma_2 \bar\gamma_3}G_{+\bar\gamma_1 \bar\gamma_2 \bar\gamma_3} &=& \bar g (\partial_\alpha \Phi - \frac{1}{2} H_{-+\alpha} - \Omega^\gamma{}_{,\gamma\alpha})\notag\\
&& -\frac{g}{2} \epsilon_\alpha{}^{\bar\gamma_1 \bar\gamma_2 \bar\gamma_3} ( \frac{1}{6}  H_{\bar\gamma_1 \bar\gamma_2 \bar\gamma_3}+  \Omega_{\bar\gamma_1, \bar\gamma_2 \bar\gamma_3}) ~.
\ey

As in the scalar case, it is convenient to re-express the first four equations in the $05$ frame
to find
\by
H_{05\alpha} &=& \Omega_{0,0\alpha} - \Omega_{5,5\alpha}~,
    \label{su4sol:H05ho}\\
\Omega_{5,0\alpha} &=& -\Omega_{\alpha,05}\label{su4sol:Omega50ho}~,\\
\Omega_{5,0\alpha} &=&
    2\Omega_{0,5\alpha} + \Omega_{\alpha,05}\label{su4sol:LieOmega}~,\\
\partial_\alpha \log f &=&
    -\tfrac{1}{2}\,\Omega_{0,0\alpha}~.\label{su4sol:Liedf}
\ey
Note that \eqref{su4sol:LieOmega} and \eqref{su4sol:Liedf} follow
directly from $\Liederivative{\KillingK}\epsilon = 0$; and
\eqref{su4sol:Omega50ho} and \eqref{su4sol:Liedf} follow from
the Killing condition $\nabla_{(A} \KillingK_{B)} = 0$.

\subsection*{(0, 2) representation}

The conditions that lie in the skew-symmetric product of two fundamental $\SU(4)$ representations
are
\by
\Omega_{[\bar \alpha, \bar\beta] -} -\Omega_{[\bar \alpha, \bar\beta] +} +\Omega_{-,\bar\alpha \bar\beta} -\Omega_{+,\bar\alpha \bar\beta} = 0~,\\
H_{-\bar\alpha \bar\beta}-H_{+\bar\alpha \bar\beta} +2 \Omega_{[\bar\alpha,\bar\beta] -}  +2 \Omega_{[\bar\alpha,\bar\beta] +} = 0 ~,\\
 \frac{1}{\sqrt{2}}f( F_{\bar\alpha \bar\beta}-G_{-+\bar\alpha \bar\beta}) + ( P^+ - P^-  ) \left(  g( H_{-\bar\alpha\bar \beta} + 2\Omega_{[\bar\alpha,\bar\beta] -}) \right)  =0~,\\
P^-\left( H_{-\bar\alpha \bar\beta} -2 \Omega_{-,\bar\alpha \bar\beta}  \right) = 0~,\\
P^-\left( g\left( H_{+\bar\alpha \bar\beta} +2 \Omega_{+,\bar\alpha \bar\beta} \right) \right) = 0~,\\
P^-\left( f( F_{\bar\alpha \bar\beta}- G_{\bar\alpha \bar\beta\gamma}{}^\gamma+G_{-+\bar\alpha \bar\beta}) +4\sqrt{2} g \Omega_{[\bar\alpha, \bar\beta] -}  \right) = 0~,\\
P^-\left( g( F_{\bar\alpha \bar\beta}+ G_{\bar\alpha \bar\beta\gamma}{}^\gamma+G_{-+\bar\alpha \bar\beta}) -4\sqrt{2} f \Omega_{[\bar\alpha, \bar\beta] +}  \right) = 0~,
\ey
where  the $P^\pm$  projectors are defined as follows
\by
 P^\pm(g\,G_{\bar\alpha\bar\beta}) &\equiv&
 \frac{1}{2} \left( g\,G_{\bar\alpha\bar\beta} \pm
     \tfrac{1}{2}\,\bar{g}\,G^{\bar\gamma_1 \bar\gamma_2}
    \epsilon_{\bar\gamma_1 \bar\gamma_2 \bar\alpha \bar\beta}\right)~,
\label{Pminus}
\ey
on any (0,2) tensor $G$. In the $05$ frame the geometric condition above can be rewritten as
\by
\Omega_{0,\bar\alpha\bar\beta} + \Omega_{[\bar\alpha,\bar\beta]0} &=& 0
    \label{su4sol:ahoahoOmega0} ~.
\ey
One can easily see that this follows from $\Liederivative{\KillingK}
\epsilon = 0$.

\subsection*{Symmetric product of fundamental representation}

The conditions that lie in the symmetric product of two fundamental $\SU(4)$ representations
are
\by
\Omega_{(\alpha,\beta)+} &=& \Omega_{(\alpha,\beta)-}~,
    \label{su4sol:Omegasymhohoplus}\\
g\,\tensor{\epsilon}{_(_\alpha^{\bar\gamma_1}^{\bar\gamma_2}^{\bar\gamma_3}}
\tensor{G}{_\beta_)_{\bar\gamma_1}_{\bar\gamma_2}_{\bar\gamma_3}} &=&
    6\sqrt{2}\,f\,\Omega_{(\alpha,\beta)+}~.
\ey
These conditions can be re-expressed in the $05$ frame as
\by
\Omega_{(\alpha,\beta)0} &=&0~,\label{su4sol:Omegasymhoho0}\\
g\,\tensor{\epsilon}{_(_\alpha^{\bar\gamma_1}^{\bar\gamma_2}^{\bar\gamma_3}}
\tensor{G}{_\beta_)_{\bar\gamma_1}_{\bar\gamma_2}_{\bar\gamma_3}} &=&
    6f\,\Omega_{(\alpha,\beta)5}~.
\label{su4sol:Gsym}
\ey
We note that geometric condition \eqref{su4sol:Omegasymhoho0} follows from the
Killing condition $\nabla\!_{(\alpha} \KillingK_{\beta)} = 0$.

\subsection*{Traceless (1,1) representation}

Given a (1,1) tensor $G_{\alpha\bar\beta}$ it can be decomposed in a (Hermitian) traceless
$\trlstens{G}{_{\alpha\bar\beta}}$
part and a trace part as $
{G}_{\alpha\bar\beta} \equiv
\tilde{G}{_{\alpha\bar\beta}}
+\frac{1}{4}\,\metric_{\alpha\bar\beta} \tensor{G}{_\gamma^\gamma}$.
Using this notation, the conditions on the fields and geometry can be written as
\by
 \trlstens{\Omega}{_\alpha_,_+_{\bar\beta}}
+\trlstens{\Omega}{_{\bar\beta}_,_+_\alpha}
&=&
 \trlstens{\Omega}{_\alpha_,_-_{\bar\beta}}
+\trlstens{\Omega}{_{\bar\beta}_,_-_\alpha}~,
\label{Killinghoahotrless}
\\
 \trlstens{H}{_-_\alpha_{\bar\beta}}
-\trlstens{H}{_+_\alpha_{\bar\beta}}
+2\trlstens{\Omega}{_[_\alpha_,_{\bar\beta}_]_-}
+2\trlstens{\Omega}{_[_\alpha_,_{\bar\beta}_]_+}
&=& 0~,
\\
f\left(
    \trlstens{G}{_\alpha_{\bar\beta}}
    -\trlstens{G}{_\alpha_{\bar\beta}_{\gamma}^{\gamma}}
    -\trlstens{G}{_-_+_\alpha_{\bar\beta}}
\right)
&=&
\sqrt{2}\,g\left(
\trlstens{H}{_-_{\bar\beta}_\alpha}
+ 2\,\trlstens{\Omega}{_\alpha_{-}_{\bar\beta}}\right)~.
\label{hoahoGHOmega}
\ey
Taking the real and imaginary parts of
\eqref{hoahoGHOmega} in the $05$ frame, the above conditions can be re-expressed as
\by
\trlstens{\Omega}{_(_\alpha_,_{\bar\beta}_)_0} &=& 0~,
    \label{su4sol:Omegasymhoaho0}\\
\trlstens{H}{_0_\alpha_{\bar\beta}} &=&
    2\trlstens{\Omega}{_[_\alpha_,_{\bar\beta}_]_5}~,
    \label{su4sol:H0hoaho}\\
\trlstens{G}{_\alpha_{\bar\beta}_\gamma^\gamma} &=&
    2\,\frac{g_1}{f}\,\trlstens{\Omega}{_(_\alpha_,_{\bar\beta}_)_5}
    +\frac{\iu\,g_2}{f}\left(
        \trlstens{H}{_5_\alpha_{\bar\beta}}
        -2\trlstens{\Omega}{_[_\alpha_,_{\bar\beta}_]_0}
    \right)~, \label{su4sol:G15}\\
\trlstens{G}{_\alpha_{\bar\beta}} + \trlstens{G}{_0_5_\alpha_{\bar\beta}} &=&
    -\frac{g_1}{f}\left(
        \trlstens{H}{_5_\alpha_{\bar\beta}}
        -2\trlstens{\Omega}{_[_\alpha_,_{\bar\beta}_]_0}
    \right)
    -\frac{(g_1 - f)(g_1 + f)}{\iu\,g_2\,f}\,
        \trlstens{\Omega}{_(_\alpha_,_{\bar\beta}_)_5}~.
\ey
The geometric condition \eqref{su4sol:Omegasymhoaho0} follows from the Killing
condition $\nabla\!_{(\alpha} \KillingK_{\bar\beta)} = 0$.

\subsection*{Traceless (1, 2) representation}

Every (1,2) tensor can be decomposed into a traceless component and a trace one as
$G_{\alpha\bar\beta\bar\gamma}=\tilde G_{\alpha\bar\beta\bar\gamma}+\frac{2}{3} g_{\alpha[\bar\beta}
G_{\bar\gamma]\delta}{}^\delta$, where $\tilde G$ denotes the traceless component. Using
this, the conditions that transform under the traceless $(1,2)$ representation are
\by
\sqrt{2}\,g\,\trlstens{G}{_{-}_\alpha_{\bar\beta_1}_{\bar\beta_2}}=f\,\trlstens{H}{_\alpha_{\bar\beta_1}_{\bar\beta_2}}
- 2 f\,\trlstens{\Omega}{_\alpha_{\bar\beta_1}_{\bar\beta_2}}
+ \frac{2}{3}\,f\left(
    \tensor{\epsilon}{_{\bar\beta_1}_{\bar\beta_2}_{\bar\gamma_1}_{\bar\gamma_2}}
    \tensor{\Omega}{_\alpha^{\bar\gamma_1}^{\bar\gamma_2}}
   -\tensor{\epsilon}{_{\bar\beta_1}_{\bar\beta_2}_{\bar\gamma_1}_{\bar\gamma_2}}
    \tensor{\Omega}{^{\bar\gamma_1}^{\bar\gamma_2}_\alpha}
\right)~,
\label{g478}
\ey
\by
\sqrt{2}\,f\,\trlstens{G}{_{+}_\alpha_{\bar\beta_1}_{\bar\beta_2}}=-g\,\trlstens{H}{_\alpha_{\bar\beta_1}_{\bar\beta_2}}
- 2 g\,\trlstens{\Omega}{_\alpha_{\bar\beta_1}_{\bar\beta_2}}
- \frac{2}{3}{\bar g}\left(
    \tensor{\epsilon}{_{\bar\beta_1}_{\bar\beta_2}_{\bar\gamma_1}_{\bar\gamma_2}}
    \tensor{\Omega}{^{\bar\gamma_1}^{\bar\gamma_2}_\alpha}
   -\tensor{\epsilon}{_{\bar\beta_1}_{\bar\beta_2}_{\bar\gamma_1}_{\bar\gamma_2}}
    \tensor{\Omega}{_\alpha^{\bar\gamma_1}^{\bar\gamma_2}}
\right)~.
\label{g479}
\ey
This concludes the description of all conditions on the fields and geometry that arise
from the Killing spinor equations.

\section{Solution of the linear system for special \texorpdfstring{$\SU(4)$}{SU(4)} backgrounds}
\label{su4solzero}

The Killing spinor \eqref{su4spinor} for special $\SU(4)$ backgrounds is simplified further as  $g_1=0$ and thus $g = \iu g_2$. Imposing  $f^2 = g \bar g$, one concludes that
$f = \pm g_2$. To derive  the conditions  on the fields and geometry, we have chosen   $f=g_2$.

\subsection*{Scalar representation}

Imposing  $f=g_2$ on the scalar conditions of the previous section, one finds that
\by
\partial_+ \log f &=& \partial_- \log f~,\\
\partial_+ \log f &=& \tfrac{1}{2}\Omega_{+,-+}~,\\
\Omega_{-,-+} &=& -\Omega_{+,-+}~,\\
\tensor{\Omega}{_-_,_\gamma^\gamma}
    +\tfrac{1}{2}\tensor{\Omega}{^\gamma_,_-_\gamma}
    -\tfrac{1}{2}\tensor{\Omega}{_\gamma_,_-^\gamma} &=& 0~,\\
       \tensor{\Omega}{_+_,_\gamma^\gamma}
        +\tfrac{1}{2}\tensor{\Omega}{^\gamma_,_+_\gamma}
        -\tfrac{1}{2}\tensor{\Omega}{_\gamma_,_+^\gamma} &=& 0
        ~,\\
\rd \Phi_+ &=& \rd \Phi_-~,\\
\rd \Phi_+ &=& -\tfrac{1}{2}\left(
        \tensor{\Omega}{^\gamma_,_+_\gamma}
        +\tensor{\Omega}{_\gamma_,_+^\gamma}
    \right) + \Omega_{+,-+}~,\\
\tensor{\Omega}{^\gamma_,_-_\gamma}
        +\tensor{\Omega}{_\gamma_,_-^\gamma} &=&
    \tensor{\Omega}{^\gamma_,_+_\gamma}
        +\tensor{\Omega}{_\gamma_,_+^\gamma} ~.
\ey
Also there is some simplification of the conditions on the $G$ and $F$ fluxes which can be expressed
as
\by
F_{-+} &=& S~,\\
H_{-\gamma}{}^\gamma &=& 2 \Omega_{-,\gamma}{}^\gamma ~,\\
H_{+\gamma}{}^\gamma &=& -2 \Omega_{+,\gamma}{}^\gamma~,\\
G_{-+\gamma}{}^\gamma &=& F_\gamma{}^\gamma + \sqrt{2} \iu (\Omega^\gamma{}_{,+ \gamma}+\Omega_{\gamma,+}{}^\gamma)~,\\
G_\gamma{}^\gamma{}_\delta{}^\delta &=& 4\sqrt{2}\iu (\Omega_{+,\gamma}{}^\gamma-\Omega_{-,\gamma}{}^\gamma) - 4 S ~,\\
\frac{\iu}{4!}(\epsilon^{\gamma_1 \gamma_2 \gamma_3 \gamma_4}G_{\gamma_1 \gamma_2 \gamma_3 \gamma_4}+\epsilon^{\bar\gamma_1 \bar\gamma_2 \bar\gamma_3 \bar\gamma_4}G_{\bar\gamma_1 \bar\gamma_2 \bar\gamma_3 \bar\gamma_4}) &=& \sqrt{2} (\Omega_{-,\gamma}{}^\gamma+\Omega_{+,\gamma}{}^\gamma)~,\\
\frac{\iu}{4!}(\epsilon^{\gamma_1 \gamma_2 \gamma_3 \gamma_4}G_{\gamma_1 \gamma_2 \gamma_3 \gamma_4}-\epsilon^{\bar\gamma_1 \bar\gamma_2 \bar\gamma_3 \bar\gamma_4}G_{\bar\gamma_1 \bar\gamma_2 \bar\gamma_3 \bar\gamma_4}) &=& \iu F_{\gamma}{}^\gamma- \frac{1}{\sqrt{2}}(\Omega^\gamma{}_{,+ \gamma}+\Omega_{\gamma,+}{}^\gamma)\\ && {} + 2\sqrt{2} \Omega_{+,-+}~.\notag
\ey
It may appear that there is an additional geometric constraint in this special case. However,
this is not the case. The number of geometric constraints is the same as for the generic
backgrounds.

\subsection*{Fundamental representation}
The conditions in the fundamental representation can be expressed as
\by
H_{-+\alpha} &=& 2 \Omega_{-,+\alpha} ~,  \\
\Omega_{+,-\alpha} &=& \Omega_{-,+\alpha} ~, \\
\Omega_{\alpha,-+} &=& -\frac{1}{2}\Omega_{-,-\alpha} + \frac{1}{2} \Omega_{+,+\alpha} ~, \\
\partial_\alpha \log f &=& \frac{1}{2} \partial_\alpha \log g \bar g = \frac{1}{4}H_{-+\alpha} -\frac{1}{2}\Omega_{\alpha,-+}-\frac{1}{2}\Omega_{-,-\alpha} ~, \\
0 &=& -2 \epsilon_\alpha{}^{\bar\gamma_1 \bar\gamma_2 \bar\gamma_3}\Omega_{\bar\gamma_1, \bar\gamma_2 \bar\gamma_3} - H_{\alpha\gamma}{}^\gamma
-2 \Omega_{\alpha,- +} ~,
\label{h1tr}
\ey
\by
\partial_\alpha \Phi &=& \frac{1}{2} H_{-+\alpha} +\frac{1}{4} H_{\alpha\gamma}{}^\gamma -\frac{1}{12}\epsilon_\alpha{}^{\bar\gamma_1 \bar\gamma_2 \bar\gamma_3}H_{\bar\gamma_1 \bar\gamma_2 \bar\gamma_3}-\frac{1}{2}\Omega_{-,-\alpha} \notag\\&&-\frac{1}{2} \Omega_{\alpha,\gamma}{}^\gamma
 + \Omega^{\gamma}{}_{,\gamma \alpha}  +\frac{1}{2}\epsilon_\alpha{}^{\bar\gamma_1 \bar\gamma_2 \bar\gamma_3}\Omega_{\bar\gamma_1, \bar\gamma_2 \bar\gamma_3} ~,
 \label{h30}
 \\
\iu(G_{-\alpha} + G_{-\alpha\gamma}{}^\gamma) &=& \frac{1}{\sqrt{2}} (H_{\alpha\gamma}{}^\gamma -2 \Omega_{\alpha,\gamma}{}^\gamma +2 \Omega_{-,-\alpha}) ~,
\label{d17}\\
\frac{\iu}{3}\epsilon_\alpha{}^{\bar\gamma_1 \bar\gamma_2 \bar\gamma_3}G_{-\bar\gamma_1 \bar\gamma_2 \bar\gamma_3} &=& \frac{1}{\sqrt{2}} (H_{\alpha\gamma}{}^\gamma -2 \Omega_{\alpha,\gamma}{}^\gamma -2 \Omega_{-,-\alpha}) ~,
\label{d18}
\\
\frac{1}{2} (3 G_{-\alpha} - G_{-\alpha\gamma}{}^\gamma) &=& \frac{\iu}{\sqrt{2}} (H_{\alpha\gamma}{}^\gamma - \frac{1}{2} \epsilon_\alpha{}^{\bar\gamma_1 \bar\gamma_2 \bar\gamma_3} H_{\bar\gamma_1 \bar\gamma_2 \bar\gamma_3}+ 2 \Omega^{ \gamma}{}_{,\gamma\alpha} \notag\\
&& \qquad+ \epsilon_\alpha{}^{\bar\gamma_1 \bar\gamma_2 \bar\gamma_3}\Omega_{\bar\gamma_1, \bar\gamma_2 \bar\gamma_3}) ~~ \\
\frac{1}{2} (3 G_{+\alpha} + G_{+\alpha\gamma}{}^\gamma) &=& \frac{\iu}{\sqrt{2}} (-H_{\alpha\gamma}{}^\gamma - \frac{1}{2} \epsilon_\alpha{}^{\bar\gamma_1 \bar\gamma_2 \bar\gamma_3} H_{\bar\gamma_1 \bar\gamma_2 \bar\gamma_3}+ 2\Omega^{ \gamma}{}_{,\gamma\alpha}\notag\\
&& \qquad - \epsilon_\alpha{}^{\bar\gamma_1 \bar\gamma_2 \bar\gamma_3}\Omega_{\bar\gamma_1, \bar\gamma_2 \bar\gamma_3}) ~, \\
\frac{\iu}{3} \epsilon_\alpha{}^{\bar\gamma_1 \bar\gamma_2 \bar\gamma_3}G_{+\bar\gamma_1 \bar\gamma_2 \bar\gamma_3}  &=& \iu (G_{+\alpha} - G_{+\alpha \gamma}{}^\gamma) - 2 \sqrt{2} \Omega_{+,+\alpha} ~,\\
\frac{1}{6\sqrt{2}} \epsilon_\alpha{}^{\bar\gamma_1 \bar\gamma_2 \bar\gamma_3}G_{+\bar\gamma_1 \bar\gamma_2 \bar\gamma_3} &=& -\iu (\partial_\alpha \Phi - \frac{1}{2} H_{-+\alpha} - \Omega^\gamma{}_{,\gamma\alpha})\notag\\
&& -\frac{\iu}{2} \epsilon_\alpha{}^{\bar\gamma_1 \bar\gamma_2 \bar\gamma_3} ( \frac{1}{6}  H_{\bar\gamma_1 \bar\gamma_2 \bar\gamma_3}+  \Omega_{\bar\gamma_1, \bar\gamma_2 \bar\gamma_3}) ~.
\label{d22}
\ey
Note that a  simplification arises because the
derivative of $g/\bar g$ vanishes.

\subsection*{(0, 2) representation}

These conditions can now be rewritten as
\by
\Omega_{[\bar \alpha, \bar\beta] -} -\Omega_{[\bar \alpha, \bar\beta] +} +\Omega_{-,\bar\alpha \bar\beta} -\Omega_{+,\bar\alpha \bar\beta} &=& 0~,\label{twoformfirst}\\
H_{-\bar\alpha \bar\beta}-H_{+\bar\alpha \bar\beta} +2 \Omega_{[\bar\alpha,\bar\beta] -}  +2 \Omega_{[\bar\alpha,\bar\beta] +} &=& 0 ~,\\
F_{\bar\alpha \bar \beta}-G_{-+\bar\alpha \bar \beta}+ 2\sqrt{2} i (\Omega_{+,\bar\alpha \bar \beta}+\Omega_{[\bar\alpha , \bar \beta] +})&=& 0 ~,\\
P^-\left( H_{-\bar\alpha \bar\beta} -2 \Omega_{-,\bar\alpha \bar\beta}  \right) &=& 0~,\\
P^+\left( H_{+\bar\alpha \bar\beta} +2 \Omega_{+,\bar\alpha \bar\beta} \right) &=& 0~,\\
P^-\left(  F_{\bar\alpha \bar\beta}- G_{\bar\alpha \bar\beta\gamma}{}^\gamma+G_{-+\bar\alpha \bar\beta}+4\sqrt{2}\,\iu\,\Omega_{[\bar\alpha, \bar\beta] -}  \right) &=& 0~,\label{ppg21a}\\
P^+\left(  F_{\bar\alpha \bar\beta}+ G_{\bar\alpha \bar\beta\gamma}{}^\gamma+G_{-+\bar\alpha \bar\beta} +4\sqrt{2}\,\iu\,\Omega_{[\bar\alpha, \bar\beta] +}  \right) &=& 0~,
\label{ppg21b}\label{twoformlast}
\ey
where $P^\pm$ is defined in \eqref{Pminus}.

It is the simplification of these equations which allows for a rather simple solution of the linear system. In particular,  two equations are automatically satisfied because  $g_1 =0$ and the remaining
conditions are simplified.

\subsection*{Symmetric product of fundamental representation}
It is straightforward to see that the conditions now become
\by
\Omega_{(\alpha,\beta)+}
&=& \Omega_{(\alpha,\beta)-}~,\\
\frac{1}{3!}
\tensor{\epsilon}{_(_\alpha^{\bar \gamma_1}^{\bar \gamma_2}^{\bar \gamma_3}}
G_{\beta)\bar\gamma_1\bar\gamma_2\bar\gamma_3}
&=& -\iu \sqrt{2} \Omega_{(\alpha,\beta)-}~.
\label{g13sym}
\ey

\subsection*{(1, 1) traceless representation}

The conditions now read as
\by
 \trlstens{\Omega}{_\alpha_,_+_{\bar\beta}}
+\trlstens{\Omega}{_{\bar\beta}_,_+_\alpha}
&=&
 \trlstens{\Omega}{_\alpha_,_-_{\bar\beta}}
+\trlstens{\Omega}{_{\bar\beta}_,_-_\alpha}~,
\label{Killinghoahotrless2}
\\
 \trlstens{H}{_-_\alpha_{\bar\beta}}
-\trlstens{H}{_+_\alpha_{\bar\beta}}
+2\trlstens{\Omega}{_[_\alpha_,_{\bar\beta}_]_-}
+2\trlstens{\Omega}{_[_\alpha_,_{\bar\beta}_]_+}
&=& 0~,
\\
\trlstens{G}{_-_+_{\bar\beta}_\alpha}
+\trlstens{G}{_{\bar\beta}^{\bar\gamma}_{\bar\gamma}_\alpha}
-\trlstens{G}{_{\bar\beta}_\alpha}
&=&
\iu\sqrt{2}\left(
\trlstens{H}{_-_{\bar\beta}_\alpha}
+ 2\,\trlstens{\Omega}{_\alpha_{-}_{\bar\beta}}\right)~.
\label{g11trl}
\ey

\subsection*{(1, 2) traceless representation}
Finally, the conditions of this representation simplify as
\by
\trlstens{H}{_\alpha_{\bar\beta_1}_{\bar\beta_2}}
- 2 \trlstens{\Omega}{_\alpha_{\bar\beta_1}_{\bar\beta_2}}
+ \frac{2}{3}\left(
    \tensor{\epsilon}{_{\bar\beta_1}_{\bar\beta_2}_{\bar\gamma_1}_{\bar\gamma_2}}
    \tensor{\Omega}{_\alpha^{\bar\gamma_1}^{\bar\gamma_2}}
   -\tensor{\epsilon}{_{\bar\beta_1}_{\bar\beta_2}_{\bar\gamma_1}_{\bar\gamma_2}}
    \tensor{\Omega}{^{\bar\gamma_1}^{\bar\gamma_2}_\alpha}
\right)
&=& \iu\sqrt{2}\,\trlstens{G}{_{-}_\alpha_{\bar\beta_1}_{\bar\beta_2}}
\label{g-12}
\\
\iu\,\trlstens{H}{_\alpha_{\bar\beta_1}_{\bar\beta_2}}
+ 2 \iu\,\trlstens{\Omega}{_\alpha_{\bar\beta_1}_{\bar\beta_2}}
-\iu \frac{2}{3}\left(
    \tensor{\epsilon}{_{\bar\beta_1}_{\bar\beta_2}_{\bar\gamma_1}_{\bar\gamma_2}}
    \tensor{\Omega}{^{\bar\gamma_1}^{\bar\gamma_2}_\alpha}
   -\tensor{\epsilon}{_{\bar\beta_1}_{\bar\beta_2}_{\bar\gamma_1}_{\bar\gamma_2}}
    \tensor{\Omega}{_\alpha^{\bar\gamma_1}^{\bar\gamma_2}}
\right)
&=& -\sqrt{2}\,\trlstens{G}{_{+}_\alpha_{\bar\beta_1}_{\bar\beta_2}}
\label{g+12}
\ey
As we shall see all the above conditions can now be solved to express the
fluxes in terms of the geometry.
\subsection{Fluxes in terms of geometry}

To continue let us decompose the fluxes as in (\ref{dec}). Furthermore, these can be decomposed in $\SU(4)$ representations. In particular for the 4-form field
strength, we have
\begin{eqnarray}
G_{(4)}&=&[G^{(4,0)}]+ [G^{(3,1)}]+ G^{(2,2)}~,~~
G_{\pm (3)}=[G_\pm^{(3,0)}]+ [G_\pm^{(2,1)}]~,~~
\cr
G_{+-(2)}&=&[G_{+-}^{(2,0)}]+ G_{+-}^{(1,1)}~,
\end{eqnarray}
where the square brackets denote the addition of the complex conjugate of the term included in the bracket. These representations can be further decomposed taking the
hermitian traces as we have already explained.

First, we find from (\ref{g-12}) and (\ref{g+12}) that
\by
\trlstens{G}{_{-}_\alpha_{\bar\beta_1}_{\bar\beta_2}}&=& \frac{\iu}{\sqrt{2}}\Big(-
\trlstens{H}{_\alpha_{\bar\beta_1}_{\bar\beta_2}}
+ 2 \trlstens{\Omega}{_\alpha_{\bar\beta_1}_{\bar\beta_2}}
- \frac{2}{3}\left(
    \tensor{\epsilon}{_{\bar\beta_1}_{\bar\beta_2}_{\bar\gamma_1}_{\bar\gamma_2}}
    \tensor{\Omega}{_\alpha^{\bar\gamma_1}^{\bar\gamma_2}}
   -\tensor{\epsilon}{_{\bar\beta_1}_{\bar\beta_2}_{\bar\gamma_1}_{\bar\gamma_2}}
    \tensor{\Omega}{^{\bar\gamma_1}^{\bar\gamma_2}_\alpha}
\right)\Big)~
\\
\trlstens{G}{_{+}_\alpha_{\bar\beta_1}_{\bar\beta_2}}&=&\frac{\iu}{\sqrt{2}}\Big(
-\trlstens{H}{_\alpha_{\bar\beta_1}_{\bar\beta_2}}
- 2 \trlstens{\Omega}{_\alpha_{\bar\beta_1}_{\bar\beta_2}}
+\frac{2}{3}\left(
    \tensor{\epsilon}{_{\bar\beta_1}_{\bar\beta_2}_{\bar\gamma_1}_{\bar\gamma_2}}
    \tensor{\Omega}{^{\bar\gamma_1}^{\bar\gamma_2}_\alpha}
   -\tensor{\epsilon}{_{\bar\beta_1}_{\bar\beta_2}_{\bar\gamma_1}_{\bar\gamma_2}}
    \tensor{\Omega}{_\alpha^{\bar\gamma_1}^{\bar\gamma_2}}\right)\Big)
\ey

Noting that
\begin{eqnarray}
G_{\alpha\bar\beta\gamma\bar\delta}&=&\tilde G_{\alpha\bar\beta\gamma\bar\delta}+\frac{1}{2} (g_{\alpha\bar\beta} \tilde G_{\gamma\bar\delta\epsilon}{}^\epsilon-
g_{\alpha\bar\delta} \tilde G_{\gamma\bar\beta\epsilon}{}^\epsilon- g_{\gamma\bar\beta} \tilde G_{\alpha\bar\delta\epsilon}{}^\epsilon+g_{\gamma\bar\delta} \tilde G_{\alpha\bar\beta\epsilon}{}^\epsilon)
\cr
&&+\frac{1}{12} (g_{\alpha\bar\beta} g_{\gamma\bar\delta}-g_{\gamma\bar\beta} g_{\alpha\bar\delta}) G_\epsilon{}^\epsilon{}_\zeta{}^\zeta~,
\end{eqnarray}
 (\ref{g11trl}) gives
\begin{equation}
\tilde G_{\alpha\bar\beta\gamma}{}^\gamma =
    \sqrt{2}\,\iu\,\big(\tilde H_{-\alpha\bar\beta}+2 \tilde\Omega_{[\alpha,\bar\beta]-}\big)
\end{equation}
\begin{equation}
\tilde G_{-+\alpha\bar\beta} =
    \tilde F_{\alpha\bar\beta} + 2\sqrt{2}\,\iu\,\tilde\Omega_{(\alpha,\bar\beta)-}
\end{equation}
The component $\tilde G^4_{(2,2)}$ is not restricted by the KSEs as well as all the
other components of the fluxes $F$, $H$ and $G$ that appear in the right-hand-side of the
above equation.

Next, we have
\begin{eqnarray}
G_{\beta\bar\gamma_1\bar\gamma_2\bar\gamma_3}=\frac{1}{6} X_{\delta\beta} \epsilon^\delta{}_{\bar\gamma_1\bar\gamma_2\bar\gamma_3}+\frac{3}{2} G_\delta{}^\delta{}_{[\bar\gamma_1\bar\gamma_2}
g_{\bar\gamma_3]\beta}~,
\end{eqnarray}
where
\begin{equation}
X_{\alpha\beta}= \epsilon_{(\alpha}{}^{\bar\gamma_1\bar\gamma_2\bar\gamma_3} G_{\beta)\bar\gamma_1\bar\gamma_2\bar\gamma_3}=-6\iu\sqrt{2} \Omega_{(\alpha, \beta)-}~,
\end{equation}
which follows from (\ref{g13sym}).

Next from (\ref{twoformfirst}) to (\ref{twoformlast}) follows that
\by
H_{-\bar\alpha \bar\beta} &=& 2 \Omega_{-,\bar\alpha \bar\beta} -4 P^+ \left(   \Omega_{-,\bar\alpha \bar\beta} + \Omega_{[\bar\alpha,\bar\beta] -} \right) ~, \\
H_{+\bar\alpha \bar\beta} &=& 2 \Omega_{+,\bar\alpha \bar\beta} + 4 \Omega_{[\bar\alpha,\bar\beta]+} -4 P^+ \left(   \Omega_{+,\bar\alpha \bar\beta} + \Omega_{[\bar\alpha,\bar\beta] +} \right)  ~, \\
G_{-+\bar\alpha \bar \beta} &=& F_{\bar\alpha \bar \beta}+ 2\sqrt{2}\,\iu\,(\Omega_{+,\bar\alpha \bar \beta}+\Omega_{[\bar\alpha , \bar \beta] +}) ~,\\
G_{\bar\alpha \bar \beta\gamma}{}^\gamma &=& 2 (P^- - P^+) F_{\bar\alpha \bar \beta}   + 2\sqrt{2}\,\iu\,P^+ \left(   \Omega_{-,\bar\alpha \bar\beta} + 3 \Omega_{[\bar\alpha,\bar\beta] -}  \right) \\
&& - 2\sqrt{2}\,\iu\,P^- \left(   \Omega_{+,\bar\alpha \bar\beta} + 3 \Omega_{[\bar\alpha,\bar\beta] +}  \right) ~.\notag
\ey
Again the $F$ flux which appears in the right-hand-side of the last two equations above
is not restricted by the KSEs.

Next let us consider (\ref{h1tr}) and (\ref{h30}) to find
\begin{eqnarray}
H_{\alpha\gamma}{}^\gamma&=&-2 \epsilon_\alpha{}^{\bar\gamma_1\bar\gamma_2\bar\gamma_3} \Omega_{\bar\gamma_1,\bar\gamma_2\bar\gamma_3}-2 \Omega_{\alpha, -+}~,
\cr
\frac{1}{12} \epsilon_\alpha{}^{\bar\gamma_1\bar\gamma_2\bar\gamma_3} H_{\bar\gamma_1\bar\gamma_2\bar\gamma_3}&=& -\partial_\alpha\Phi+ \partial_\alpha \log f
+\frac{1}{2} \Omega_{-,+\alpha} -\frac{1}{2} \Omega_{\alpha,\gamma}{}^\gamma+ \Omega_{\bar\gamma,}{}^{\bar\gamma}{}_\alpha ~.
\end{eqnarray}

Next we shall solve (\ref{d17}) to (\ref{d22}) using the above expressions for the $H$ flux. First (\ref{d18}) gives
\begin{equation}
\frac{\iu}{3} \epsilon_\alpha{}^{\bar\gamma_1\bar\gamma_2\bar\gamma_3} G_{-\bar\gamma_1\bar\gamma_2\bar\gamma_3}=\frac{1}{\sqrt{2}} \big(-2
\epsilon_\alpha{}^{\bar\gamma_1\bar\gamma_2\bar\gamma_3} \Omega_{\bar\gamma_1,\bar\gamma_2\bar\gamma_3} -2 \Omega_{\alpha, -+} -2\Omega_{\alpha,\gamma}{}^\gamma-2 \Omega_{-,-\alpha}\big)~.
\end{equation}
Then one finds that
\begin{multline}
\tensor{G}{_-_\alpha_\gamma^\gamma} = \frac{\iu}{2\sqrt{2}}\big(
  4 \epsilon_\alpha{}^{\bar\gamma_1\bar\gamma_2\bar\gamma_3} \Omega_{\bar\gamma_1,\bar\gamma_2\bar\gamma_3}
  + 5 \Omega_{\alpha,-+}
  - 6 \partial_\alpha\Phi\\
  + 6 \partial_\alpha \log f
  + 3 \Omega_{-,+\alpha}
  - 3 \Omega_{-,-\alpha}
  + 4 \Omega_{\bar\gamma,}{}^{\bar\gamma}{}_\alpha
\big)
\end{multline}
and
\begin{equation}
F_{-\alpha} = \frac{\iu}{2\sqrt{2}}\left(
  - \Omega_{\alpha,-+}
  + 4 \Omega_{\alpha, \gamma}{}^\gamma
  - 3 \Omega_{-,+\alpha}
  - \Omega_{-,-\alpha}
  + 6 \partial_\alpha\Phi
  - 6 \partial_\alpha \log f
  - 4 \Omega_{\bar\gamma,}{}^{\bar\gamma}{}_\alpha
\right)~,
\end{equation}
and similarly
\begin{equation}
\frac{\iu}{6} \epsilon_\alpha{}^{\bar\gamma_1\bar\gamma_2\bar\gamma_3} G_{+\bar\gamma_1\bar\gamma_2\bar\gamma_3}=\frac{1}{\sqrt{2}}\big(2\partial_\alpha\log f-\Omega_{-,+\alpha}
-\Omega_{\alpha,\gamma}{}^\gamma+ \epsilon_\alpha{}^{\bar\gamma_1\bar\gamma_2\bar\gamma_3} \Omega_{\bar\gamma_1,\bar\gamma_2\bar\gamma_3}\big)~,
\end{equation}
\begin{equation}
F_{+\alpha} = \frac{\iu}{\sqrt{2}}\left(
  \Omega_{\alpha,-+}
  + 2 \Omega_{\alpha, \gamma}{}^\gamma
  - \Omega_{-,+\alpha}
  - \Omega_{+,+\alpha}
  + 3 \partial_\alpha\Phi
  - 4 \partial_\alpha \log f
  - 2 \Omega_{\bar\gamma,}{}^{\bar\gamma}{}_\alpha
\right)~,
\end{equation}
\begin{equation}
\tensor{G}{_+_\alpha_\gamma^\gamma} = \frac{\iu}{\sqrt{2}}\left(
  2 \epsilon_\alpha{}^{\bar\gamma_1\bar\gamma_2\bar\gamma_3} \Omega_{\bar\gamma_1,\bar\gamma_2\bar\gamma_3}
  + \Omega_{\alpha,-+}
  + 3 \partial_\alpha\Phi
  - 3 \Omega_{-,+\alpha}
  + 3 \Omega_{+,+\alpha}
  - 2 \Omega_{\bar\gamma,}{}^{\bar\gamma}{}_\alpha
\right)~.
\end{equation}

Furthermore the scalar conditions give

\by
F_{-+} &=& S~,\\
H_{-\gamma}{}^\gamma &=& 2 \Omega_{-,\gamma}{}^\gamma ~,\\
H_{+\gamma}{}^\gamma &=& -2 \Omega_{+,\gamma}{}^\gamma~,\\
G_{-+\gamma}{}^\gamma &=& F_\gamma{}^\gamma + \sqrt{2}\,\iu\,(\Omega^\gamma{}_{,+ \gamma}+\Omega_{\gamma,+}{}^\gamma)~,\\
G_\gamma{}^\gamma{}_\delta{}^\delta &=& 4\sqrt{2}\,\iu\,(\Omega_{+,\gamma}{}^\gamma-\Omega_{-,\gamma}{}^\gamma) - 4 S ~,\\
\frac{\iu}{4!}\epsilon^{\gamma_1 \gamma_2 \gamma_3 \gamma_4}G_{\gamma_1 \gamma_2 \gamma_3 \gamma_4} &=& \frac{\sqrt{2}}{2} (\Omega_{-,\gamma}{}^\gamma+\Omega_{+,\gamma}{}^\gamma)+
\frac{\iu}{2} F_{\gamma}{}^\gamma- \frac{1}{2\sqrt{2}}(\Omega^\gamma{}_{,+ \gamma}+\Omega_{\gamma,+}{}^\gamma)\\ && + \sqrt{2} \Omega_{+,-+}~,\notag~
\ey
Again the component of the $F$ flux and $S$ that appear in the right-hand-side of the
above equations are not restricted by the KSEs.

\end{document}